\documentclass[aps,prl,preprint,superscriptaddress]{revtex4}
\usepackage{graphicx}
\begin{document}

\title{Band dispersion in the deep $1s$ core level of graphene}
\author{S. Lizzit}
\affiliation{Sincrotrone Trieste, S.S. 14 Km. 163.5, 34149
Trieste, Italy}
\author{G. Zampieri}
\affiliation{Centro At\'{o}mico Bariloche and Instituto
Balseiro,\\
Comisi\'{o}n Nacional de Energ\'{i}a At\'{o}mica,
8400-Bariloche,\\
Argentina}
\author{L. Petaccia}
\affiliation{Sincrotrone Trieste, S.S. 14 Km. 163.5, 34149
Trieste, Italy}
\author{R. Larciprete}
\affiliation{CNR-Institute for Complex Systems, Via Fosso del Cavaliere 100, 00133 Roma, Italy}
\author{P. Lacovig}
\affiliation{Sincrotrone Trieste, S.S. 14 Km. 163.5, 34149
Trieste, Italy}
\affiliation{Physics Department, University of Trieste, Via Valerio 2, 34127 Trieste, Italy}
\author{E.~D.~L.~Rienks}
\affiliation{Institute for Storage Ring Facilities and Interdisciplinary Nanoscience Center (iNANO), University of Aarhus,
8000 Aarhus C, Denmark}
\author{A. Baraldi}
\affiliation{Physics Department, University of Trieste, Via Valerio 2, 34127 Trieste, Italy}
\affiliation{CENMAT, University of Trieste, Via Valerio 2, 34127 Trieste, Italy}
\affiliation{Laboratorio TASC INFM-CNR, AREA Science Park, S.S. 14 Km 163.5, 34149 Trieste, Italy}
\author{Ph. Hofmann}
\affiliation{Institute for Storage Ring Facilities and Interdisciplinary Nanoscience Center (iNANO), University of Aarhus,
8000 Aarhus C, Denmark}
\email[]{philip@phys.au.dk}
\homepage[]{http://www.phys.au.dk/~philip/}
\date{\today}

\begin{abstract}
Chemical bonding in molecules and solids arises from the overlap of valence electron wave functions, forming extended molecular orbitals and dispersing Bloch states, respectively. Core electrons with high binding energies, on the other hand, are localized to their respective atoms and their wave functions do not overlap significantly. Here we report the observation of band formation and considerable dispersion (up to 60 meV) in the $1s$ core level of the carbon atoms forming graphene, despite the high C $1s$ binding energy of $\approx$~284~eV. Due to a Young's double slit-like interference effect, a situation arises in which only the bonding or only the anti-bonding states is observed for a given photoemission geometry. 
 \end{abstract}

\maketitle

The assumption that electrons in deep core states do not participate in the bonding of solids, and thus do not show band-like dispersion, is at the base of our understanding of the solid state. It is also of essential practical significance for many experiments, such as x-ray photoelectron spectroscopy. This technique derives its power from the fact that the precise value of the core level binding energy depends on the chemical environment of the emitting atoms, but it is tacitly assumed that it has a single, well-defined energy, i.e. it does not show any dispersion. Violations of this assumption have only been found for small molecules in the gas phase such as C$_2$H$_2$  or N$_2$ with much stronger bonding and shorter bonding distances than present in solids  \cite{Kempgens:1997,Hergenhahn:2001}.  In this paper we report the observation of a considerable band-like dispersion of the C $1s$ core level in graphene. The dispersion is observed as an emission-angle dependent binding energy modulation and it is shown that under appropriate conditions only the bonding or anti-bonding states can be observed. 

\begin{figure}
  \includegraphics[width=3.25in]{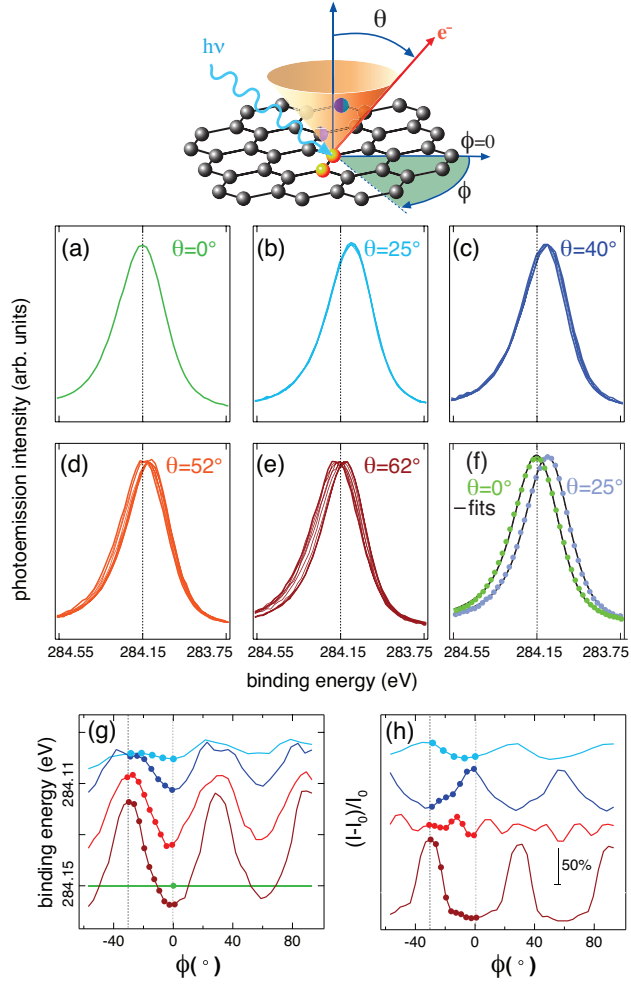}\\
  \caption{(a)-(e) C 1$s$ photoemission spectra taken at a photon energy of 400~eV, for fixed polar emission angles $\theta$ in each panel but at different azimuthal emission angles $\phi$ (see sketch of the experimental geometry at the top of the figure). The spectra are all normalized to the same height and shown as a group plot, such that binding energy variations become evident. The first and last spectrum in a range of azimuthal angles $\phi$ is indicated by a thicker line. (f) Comparison of the spectrum taken at $\theta=0^{\circ}$ and one taken at 25$^{\circ}$. The lines are the fits through the data points using the lineshape parameters described in the supplementary material. (g)  C $1s$ binding energy  required to obtain a good fit for the curves shown above (markers) as well as for the entire azimuthal range measured (lines). The green horizontal line marks the binding energy at normal emission. The binding energy uncertainty is smaller than 10~meV. Error bars are not shown.  (h)
  Intensity variation of the C~$1s$ peak as a function of azimuthal angle.  The curves are shifted vertically for clarity. }
  \label{fig:1}
\end{figure}

The binding energy modulations of the C~$1s$ core state are illustrated in Fig. \ref{fig:1}(a)-(e). Each panel shows a group of spectra taken at a fixed polar emission angle $\theta$ as a function of azimuthal emission angle $\phi$. Clear shifts of the peak position are observed. Fig. \ref{fig:1}(f) shows a comparison of the spectrum taken at normal emission and one taken at $\theta=25^{\circ}$ together with the result of a peak fit to these two spectra. Details of the fitting procedure are described in the supporting material.

 The binding energy variation obtained from the peak fitting is given in Fig. \ref{fig:1}(g)  with the markers corresponding to the spectra in (a)-(e). Strong changes are evident with the largest difference of binding energies spanning $\approx$60~meV. The variation is consistent with the point symmetry of the graphene lattice. Fig. \ref{fig:1}(h) shows the 
 intensity variations of the peaks, displayed as the modulation function (see supplementary material). Strong intensity modulations are observed, caused by photoelectron diffraction in the final state. The variations follow the point symmetry of the graphene lattice, but they do not appear to be correlated with the binding energy in (g), with the main structures being in phase for some polar emission angles and out of phase for others. 

\begin{figure}
 \includegraphics[width=3.25in]{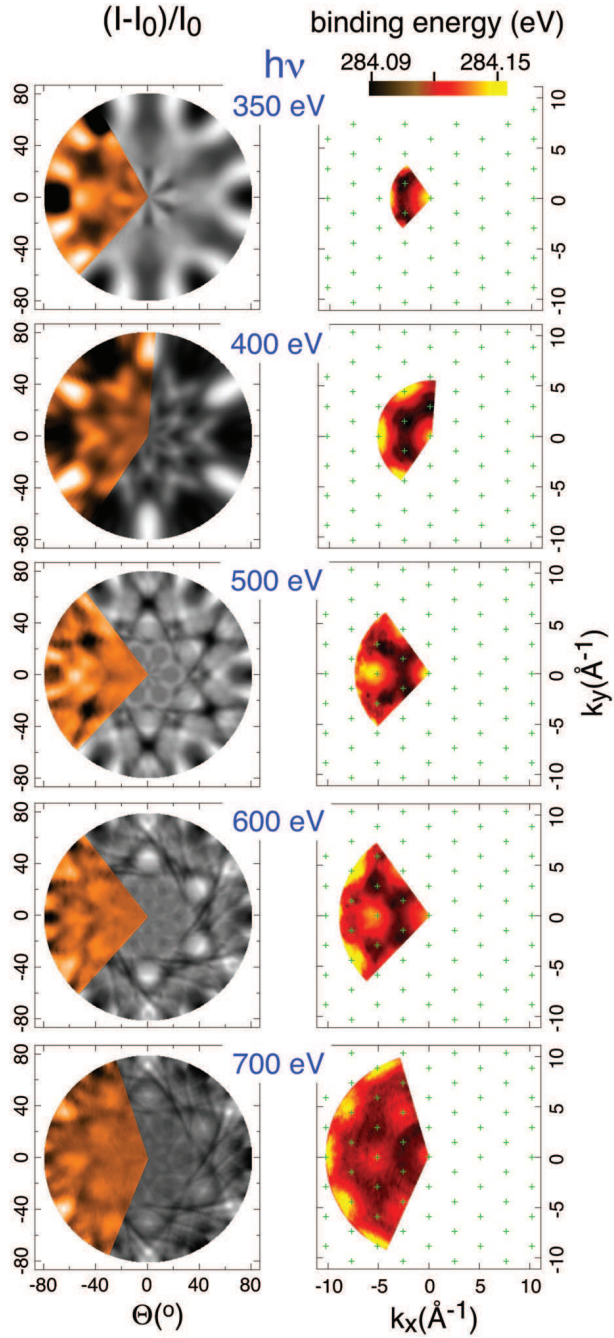}\\
  \caption{Left panel: stereographic projection of the photoemission intensity modulation as a function of emission angle for scans taken at different photon energies $h\nu$. The colored fraction of the disk are the data, the greyscale part is a calculation of the expected intensity. Right panel: The corresponding binding energy variations obtained from the peak fitting. Note that these variations are shown as a function of the wave vector parallel to the surface rather than emission angle. The binding energy uncertainty is of the order of 10~meV. The green crosses correspond to the reciprocal lattice points of graphene, \emph{i.e.} to the $\bar{\Gamma}$ points. }
  \label{fig:2}
\end{figure}

While the data of Fig. \ref{fig:1} serve to illustrate the nature of the effect and the fitting procedure, they are insufficient to pin down the physical origin of the modulation. Fig. \ref{fig:2} therefore shows a much more extensive data set measured over many polar and azimuthal angles and at different photon energies. Again, a good fit to all the spectra in the data set was obtained for a single C $1s$ component using always the same line shape. Note that this excludes the existence of unresolved components, because their intensities would modulate differently, changing the shape of the peak. The left panel of the figure shows the resulting intensity modulation function while the right panel gives the binding energy modulation. The modulation function is compared to the simulation for a flat, free-standing layer of graphene. The agreement between experiment and calculation is excellent.

The binding energy modulation, on the other hand, is shown as a
function of $\mathbf{k}_{\parallel}$, the wave-vector component parallel to the surface, which
is the only relevant wave-vector for a two-dimensional system like
graphene. As the portion of the reciprocal space covered by the experiment
increases at higher photon energies, a periodic pattern emerges which, 
however, does not coincide with the reciprocal lattice mesh. 
At the origin ($\mathbf{k}_{\parallel}=(0,0)$) the binding energy is
always close to its maximum value. The experiments at $h\nu=400$~eV and 500~eV show that the
binding energy takes its maximum value also at the next-nearest 
neighbor reciprocal lattice points, while the
experiments at 600 and 700 eV show that this occurs again at the next-nearest
neighbors of these latter points. Interestingly, at all the other reciprocal lattice
points the binding energy takes its minimum value. The periodic pattern, therefore, is
described with two complementary sublattices: the binding energy is at its maximum
value at all the points connected by vectors that are $\sqrt{3}$ times longer than the primitive 
vectors and rotated by 30$^{\circ}$,  and it is at its minimum value at all
the other reciprocal lattice points.

There are, in principle, several different mechanisms which could lead to the observed binding energy variations. The first is the existence of several unresolved components which change their relative intensity due to a final state effect (photoelectron diffraction) and thereby mimic a peak-shift. This appears highly unlikely in the present case, not only because of the excellent agreement with the intensity calculation for a single component, but also because a peak shift of the observed magnitude would essentially require a complete supression of peak intensities in certain directions, and this is unrealistic for a usual photoelectron diffraction-type modulation. It would also be very unlikely that the mechanism leads to a binding energy modulation which is periodic \emph{in reciprocal space}, as seen in Fig. \ref{fig:2}.

The second possible explanation is a  recoil effect in which some of the photoelectron's energy is used to excite lattice vibrations. This effect has been reported for photoemission from the graphite C $1s$ state  \cite{Takata:2007}. For increasing photon energies it leads to a decreasing apparent binding energy since some of the photoelectron's energy remains with the emitting atom. The effect can be of considerable size, several hundred meV for photon energies of several keV, but its magnitude should be insignificant for the low energies used here. One would not expect it to lead to any periodic modulation either.

\begin{figure}
  \includegraphics[width=3.25in]{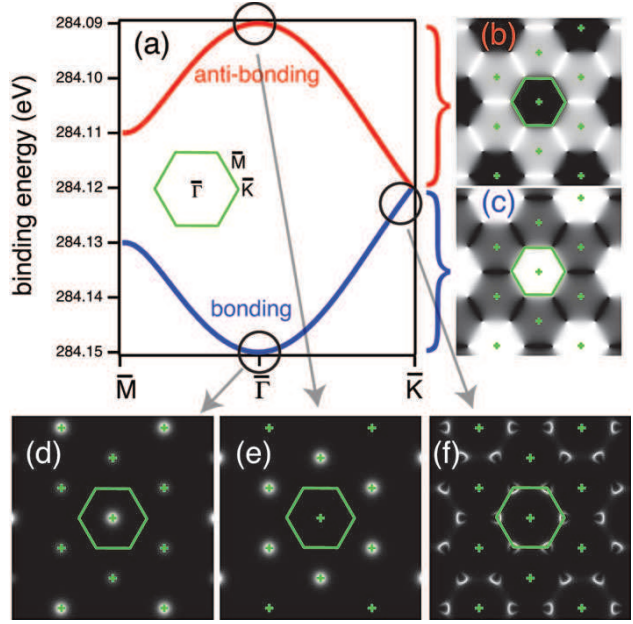}\\
  \caption{(a) Result of a tight-binding calculation for a $\sigma$-type band formed from the C~$1s$ core states in graphene. The bonding (blue) and anti-bonding (red) bands are degenerate at $\bar{K}$ and show the highest splitting at $\bar{\Gamma}$. The inset shows the Brillouin zone of graphene. (b) and (c) Calculated photoemission intensity from all the anti-bonding and bonding states, respectively. The grey-scale is chosen such that bright corresponds to high intensity. The green crosses mark the reciprocal lattice of graphene and the green hexagon the first Brillouin zone. (d)-(f) Calculated photoemission intensity from the states in the binding energy windows indicated by the small circles in (a). Note the similarity of the emission pattern from the bonding states in (d) with the positions of maximum binding energy in the right panel of Fig. \ref{fig:2}. }
  \label{fig:3}
\end{figure}

This leaves a third possibility which is an initial state effect, \emph{i. e.} a band-like dispersion of the initial state. The simplest conceivable picture for this is the formation of a $\sigma$-type band between the 1$s$ states of the two atoms in the unit cell of graphene, highlighted in orange in the sketch of Fig. \ref{fig:1}. A tight-binding calculation of such a band is shown in Fig. \ref{fig:3}(a). The absolute binding energy and the band width are arbitrarily chosen to mimic those observed here. The dispersion shows two bands with the highest energy separation at $\bar{\Gamma}$ and degeneracy at the $\bar{K}$ point of the two-dimensional Brillouin zone. In the following we refer to these bands, somewhat loosely, as the bonding and the anti-bonding band.

On the face of it, it appears hard to reconcile such a dispersion with the experimental observation, as one would expect to observe a single, narrow C $1s$ peak at  $\bar{K}$ and a broad or even split peak at $\bar{\Gamma}$. This is clearly not supported by the data. Also, the $\sigma$-band should be periodic in reciprocal space, \emph{e. g.} the peak position and width should be the same at all $\bar{\Gamma}$ points. According to the right panel of Fig. \ref{fig:2}, this is not the case either. Most of the $\bar{\Gamma}$ points marked in the figure appear close to either a maximum or a minimum in the binding energy but clearly the observed periodicity is not the same as that of the reciprocal lattice.

The hypothesis of band dispersion and the experimental data can, however, be reconciled when taking into account a curious interference effect which is caused by the presence of two atoms in the unit cell of graphene. In the most simple picture, this interference can be interpreted as a type of Young's double slit phenomenon in which the two atoms in the unit cell act as electron sources. The effect has been studied in detail for the valence band of graphite and graphene \cite{Shirley:1995b,MuchaKruczynski:2008}. 

As an illustration of the interference effect, we have calculated the expected photoemission intensities for all the anti-bonding and all the bonding states. The results are given in  Fig. \ref{fig:3}(b) and (c), respectively. The strength of the interference effect is evident: in the first Brillouin zone, for instance, emission from the anti-bonding states is entirely suppressed while it is intense from the bonding states. In the neighboring zones it is the other way round. Figs. \ref{fig:3} (d)-(f) show the emission intensity for smaller energy windows at the top of the anti-bonding band, at the bottom of the bonding band, and just below the Dirac point at  $\bar{K}$. In the last case, a constant energy contour shows a triangular shape, as expected for the non-linear dispersion away from $\bar{K}$. The photoemission intensity around this triangular contour shows strong variations which are caused by the interference effect and very similar to the results obtained for the valence $\pi$-band of graphite and graphene \cite{Shirley:1995b,MuchaKruczynski:2008}. 

As already expected from Fig. \ref{fig:3}(b) and (c), the interference effect is even stronger for emission from the bonding and anti-bonding states at $\bar{\Gamma}$ (see Fig. \ref{fig:3}(d) and (e)). The intensity variations of both bands are almost opposite to each other: for some $\bar{\Gamma}$ points only the bonding band is observed, for others only the anti-bonding band. For normal emission this is easy to understand: for the bonding band the wave functions centred on the two atoms in the unit cell emit in phase and this band is observed. For the anti-bonding wave function, the two atomic wave functions emit out of phase, thus suppressing the photoemission. 

The presence of the interference effect easily reconciles the 
 hypothesis of a $\sigma$-band formation with the data. First of all, it explains the fact that the peak shape is similar for all emission directions. For emission near $\bar{K}$ the peak is narrow because of the degeneracy of the bands at $\bar{K}$. At the $\bar{\Gamma}$ points the peak is also narrow, in contrast to naive expectation, because it does not show both the bonding and the anti-bonding bands but rather only one of them at every given  $\bar{\Gamma}$ point. The interference effect is also responsible for the apparently incorrect periodicity of the band dispersion in reciprocal space. The situation is not such that all  $\bar{\Gamma}$ points are equivalent,  because one either observes emission from the bonding or from the anti-bonding band. Note that the sign of the observed modulation is consistent with this interpretation, too. The binding energy of the peak observed at normal emission is close to the global maximum of the entire data set, consistent with emission from the bonding $\sigma$-band. In fact, the calculated emission from the bonding band shown in Fig. \ref{fig:3}(d) can be compared directly to the plot of the apparent binding energy in Fig. \ref{fig:2}, which is also scaled such that high binding energies are bright. 
 
The size of the bonding / anti-bonding splitting in graphene can be inferred from the difference of observed binding energies at inequivalent $\bar{\Gamma}$ points. Combing all the available data from  $\bar{\Gamma}$ points showing either the bonding or the anti-bonding band, we evaluate the size of the splitting to be $60\pm10$~meV. We can compare this value to the size of the bonding / anti-bonding splitting in small carbon-containing molecules such as C$_2$H$_2$ (C-C distance of 1.2~\AA, splitting of 105~meV \cite{Kempgens:1997}) and C$_2$H$_4$ (C-C distance of 1.34~\AA~expected splitting of 20-30~meV \cite{Koppel:1997}) if we assume that the matrix element for hopping between the core electrons on the two carbon atoms depends exponentially on the C-C distance and that the size of the splitting scales linearly with the number of nearest neighbors. Assuming a splitting of 105~meV and 25~meV for   C$_2$H$_2$ and C$_2$H$_4$, respectively, we would expect a total bandwidth of $\approx$30~meV for graphene. The observed magnitude of the dispersion is somewhat larger but in the right order of magnitude. 

Apart from the fundamental importance of our results for the bonding in solids, a dependence of the core level binding energy on the emission angle can have  implications for the interpretation of high-resolution core level data from graphene, graphite and related materials. The absolute magnitude of the binding energy variation observed here is appreciable compared with usual chemical shifts and could easily be interpreted incorrectly. In a limited data set, for instance, an apparent shift of the C~$1s$ peak as a function of polar emission angle might be mistaken for signs of a surface core level shift in graphite because of the higher surface sensitivity for off-normal emission. Ignoring dispersion effects might play some role in the recent dispute about the existence of a surface core level shift in graphite \cite{Balasubramanian:2001b,Smith:2002,Lizzit:2007,Hunt:2008}, but note that the interference effect reported here is inconsistent with the observation of multiple C~$1s$ components. Here we can state that to a good approximation either the bonding or the anti-bonding band can be observed in any given emission direction. 

In conclusion, we have shown that the interaction between the two atoms in the unit cell of graphene is sufficiently strong to induce the formation of a $\sigma$-band derived from the carbon $1s$ state. Interference effects in the photoemission process give rise to a strong suppression of  either the bonding or the anti-bonding band for different emission directions, such that the absolute size of the splitting could be determined from the entire data set. The interference effect furthermore allows to observe only the bonding or the anti-bonding states under certain conditions, opening the unique opportunity for detailed and possibly local as well as time-resolved studies of bonding in solids.

\section{Acknowledgements}
We gratefully acknowledge stimulating discussions with Gustav Bihlmayer, D.~L. Mills and  Javier Garc\'{\i}a de Abajo. The research leading to these results has received funding from the Danish National Research Council, the Lundbeck Foundation and the European Community's Seventh Framework Programme (FP7/2007-2013) under grant agreement no. 226716. GZ thanks ANPCyT of Argentina and the ICTP Trieste for financial support
under the TRIL Programme.

\section{Supporting material: Experimental and calculational methods}

 The graphene film on Ir(111) was prepared \emph{in situ} using standard procedures \cite{Pletikosic:2009} and the sample quality and cleanliness was monitored by low energy electron diffraction (LEED) and x-ray photoelectron spectroscopy. Angle-resolved photoemission data were taken at the SuperESCA beamline of the storage ring ELETTRA. The experimental chamber is equipped with a 150~mm electron energy analyser from SPECS, implemented with a  delay line detector developed in-house. It allows fast acquisition of the spectra in snap-shot mode, with up to thousand data points in each spectrum and in as short as 80~ms/spectrum. The data shown in Fig. \ref{fig:2} are acquired in scanning mode at $h\nu=$350~eV, 400~eV and 700 eV, while in snap-shot mode at 500~eV and 600~eV.  The fast scans, with a total of 420 spectra that fill 1/3 of the emission hemisphere, take less than 30 min. At the other photon energies more spectra were taken: 2270 at $h\nu=700$~eV,  820 at $h\nu=400$~eV and 1300 at $h\nu=350$~eV. Such scans take several hours. The photon energy was calibrated by the absorption edges of gas-phase Ar, Ne and N$_2$. The manipulator used in the experiments is a modified version of the VG-CTPO with fully computer-controlled polar and azimuthal rotations. 

For the analysis of the data all the spectra taken at one photon energy were first aligned to the Fermi level of the Ir substrate, then they were fitted with the same line shape parameters but the binding energy was left free in the fit. Fits for two spectra are shown in Fig. \ref{fig:1}(f). The actual peak fits were made using a Doniach-Sunjic line profile \cite{Doniach:1970} with a Lorentzian width of 130~meV, an asymmetry parameter of 0.093,  a Gaussian width depending of the photon energy and a linear background. The narrowest Gaussian width was 165~meV and this could be used for photon energies up to 400~eV. It is the value used for the two fits shown in Fig. \ref{fig:1}. Note, however, that the spectra were measured and fitted in a wider range than displayed in Fig. \ref{fig:1}. We have observed a small variation of the lineshape over the entire data set taken at any given photon energy. However, a good fit to all the spectra can be obtained with one set of parameters describing the line shape. Arbitrariness in the exact values of these parameters induces a small uncertainty in the absolute degree of binding energy modulation of about 5~meV. We thus estimate the total uncertainty of the observed bandwidth to be of the order of 10~meV.

The modulation functions  in Fig. \ref{fig:1}(h) and the left part of Fig. \ref{fig:2} were obtained for each polar emission angle $\Theta$ from the peak intensity $I(\Theta,\phi)$ as $(I(\Theta,\phi)-I_0(\Theta))/I_0(\Theta$), were $I_0(\Theta)$ is the average value of each azimuthal scan. Note that no artificial symmetry was imposed on the data. The experimental modulations are as symmetric as they appear, greatly enhancing out confidence in the data. The calculated modulations in Fig.   \ref{fig:2} were obtained in the same way. 

The potoemission intensity calculations used to obtain the calculated modulation functions in Fig. \ref{fig:2} were performed by the program package for Electron Diffraction in Atomic Clusters (EDAC) \cite{Abajo:2001}. The photoemission intensity at each emission angle was calculated as the incoherent sum of the intensities for the two atoms in the graphene unit cell. This approach might, at first sight, appear inconsistent with the main conclusion of this paper that the $1s$ state is described as an extended coherent wave function. Note, however, that what we calculate and compare to is the sum of the bonding and anti-bonding states. The sum of the probability density for the wave functions is strongly peaked at the atomic cores, independent of the wave vector chosen. Consequently, one would expect the same photoemission signal, including final-state diffraction effects, as from the sum of all core states. The simulations were performed on a flat, free-standing graphene layer. The lattice parameter was set to 2.466~\AA. The influence of the underlying Ir(111) substrate, that of small changes of the lattice parameter, as well as that of a relatively small corrugation of the graphene layer with moir\'e periodicity \cite{Diaye:2008} was tested. No significant changes were found in the diffraction patterns. Parameters for the calculations are: $V_0 = 8$~eV, linear polarization, angle between incoming light and analyzer 70$^{\circ}$, angular acceptance 10$^{\circ}$, system at room temperature. The Debye temperature of graphene is much higher than this and hence its precise value does not lead to any noticeable differences in the result. As seen in Fig. \ref{fig:2} the agreement between these calculation and the data is excellent with the two being nearly indistinguishable. However, in order to obtain this agreement, the photoelectron's kinetic energy in the simulation had to be assumed to be 12~eV lower than that in the experiment. The reason for this is presently unknown.
 
The tight-binding calculations for the band structure and photoemission intensity in Fig. \ref{fig:3} are very similar to those presented for the $\pi$-valence band in Ref. \cite{MuchaKruczynski:2008}, but use $\sigma$-band symmetry instead. The parameters used for the calculation are inspired by the experimental binding energy and band width but they are only chosen for illustrative purposes. Their precise value is irrelevant for the conclusions of this paper.

%\bibliographystyle{nature}
%\bibliography{groupreferences_new}
\end{document}